# A REVIEW OF ONE-WAY AND TWO-WAY EXPERIMENTS TO TEST THE ISOTROPY OF THE SPEED OF LIGHT

Md. Farid Ahmed[1,*], Brendan M. Quine[1,2], Stoyan Sargoytchev [3], and A. D. Stauffer[2]

[1]Department of Earth and Space Science and Space Engineering, York University, 4700 Keele Street, Toronto, Ontario, Canada-M3J 1P3.
[2]Department of Physics and Astronomy, York University, 4700 Keele Street, Toronto, Ontario, Canada-M3J 1P3.
[3]Department of Computer Science and Engineering, York University, 4700 Keele Street, Toronto, Ontario, Canada-M3J 1P3.

[*]E-mail address of corresponding author : mdfarid@yorku.ca

**Abstract** **:** As we approach the 125[th] anniversary of the Michelson-Morley experiment in 2012, we review experiments that test the isotropy of the speed of light. Previous measurements are categorized into one-way (single-trip) and two-way (round-trip averaged or over closed paths) approaches and the level of experimental verification that these experiments provide is discussed. The isotropy of the speed of light is one of the postulates of the Special Theory of Relativity (STR) and, consequently, this phenomenon has been subject to considerable experimental scrutiny. Here, we tabulate significant experiments performed since 1881 and attempt to indicate a direction for future investigation.

**Keywords** **:** Isotropy test, Special theory of relativity, Constancy of the one-way speed of light.

**PACS Nos.** **:** 03; 03.+p, 07; 07.60.-j, 07.60.Ly

## 1. Introduction

In 1905 Albert Einstein introduced the Special Theory of Relativity (STR) [1] – a theoretical framework that proved immediately successful in unifying Maxwell's electrodynamics with classical Mechanics. One of the primary experimental measures of STR was that it provided an explanation for the results of Michelson and Morley's investigation that found no variation in the speed of light with Earth motion [2, 3]. Under STR, the laws of electrodynamics, as expressed by Maxwell's equations, were held invariant under Lorentz transformations as a consequence of the assumption that the velocity of light is constant in all systems independent of the velocity of the light source. This theory did not only resolve open questions in electrodynamics, it also introduced a revolutionary new notion of space and time as a single entity, space-time. The main feature of the STR, the space-time symmetry of Local Lorentz Invariance (LLI), has influenced profoundly the development of fields from science-technology to philosophy [4 - 6]. Indeed, our present understanding of all physical theories describing nature are based on Special and General Relativity (GR) – the constancy of the speed of light being necessary for the validity of both relativity theories. LLI is required by GR in the limiting case of negligible gravitation and is today the basis of the standard model of particle physics (relativistic quantum field theory). Despite the remarkable success of STR and GR several modern theoretical approaches have begun to predict variation on the constant light-speed postulate. String theory which seeks to unify today's standard model with general relativity predicts a violation of the constancy of the speed of light [7 - 10]. Another approach has been described by Zhou and Ma who have proposed a new framework as the Standard Model Supplement (SMS) which brings new terms violating Lorentz invariance in the standard model [11]. Also, Albrecht and Magueijo have proposed the Variable Speed of Light (VSL) theory in order to explain some significant cosmological problems [12]. However, all theoretical predictions of the violation of the Lorentz Invariance are speculative which lack experimental verification.

The widely used experiments to test the STR may be divided into three classical types based on Robertson [13], and Mansouri and Sexl [14 – 16] as: (a) Michelson-Morley (M-M type) [3] which tests the isotropy of the speed of light, (b) Kennedy-Thorndike (K-T type) [17] which tests the velocity dependence of the speed of light, and (c) Ives and Stilwell (I-S type) [18] which tests the relativistic time dilation. These experiments have been reviewed previously by different authors [19 – 23]. Most of these experiments especially M-M type and K-T type only test the two-way speed of light (in a closed path of given length). However, still there are questions about the constancy of the one-way speed of light [24, 25]. Here, we present a comparison and review of the experimental tests which cover isotropy of the velocity of light: one-way and two-way speed of light measurements.

## 2. Theoretical frameworks to interpret the experiments

The first experiments were performed by Michelson in 1881 in Potsdam [2] and then in 1887 together with Morley in Cleveland [3]. These experiments intended to detect the presence of the ether-drift were interpreted based on the concept of a hypothetical inertial frame of reference. The failure of these experiments to detect an inertial frame led to the dismissal of the ether frame concept. After the discovery of the Cosmic Microwave Background (CMB) in 1965 [26], an alternate basis for a frame of reference became identified with the CMB. A brief review of the ether frame, the CMB frame and also commonly used test theories are presented in the following sections.

### 2.1. The ether frame :

The "ether frame" which was called the solar "rest frame" by Einstein is a preferred inertial reference frame in which the speed of light is isotropic and is predicted by Maxwell's equations of electrodynamics. The moving frame of reference can be moving (translating and rotating) freely of its own accord, or it can be imagined to be attached to a physical object. A kinematic quantity measured relative to the fixed inertial frame is considered absolute (e.g., absolute velocity), and those measured relative to the moving frame are termed relative (e.g., relative velocity).

The motion relative to Earth's centre of mass on the equator of the Earth is about $5 \times 10^2\ \mathrm{ms}^{-1}$. As well the Earth travels at a speed of around $3 \times 10^4\ \mathrm{ms}^{-1}$ in its orbit around the Sun. Also the Sun is traveling together with its planets about the galactic centre with a speed $2.5 \times 10^5\ ms^{-1}$, and there are other motions at higher levels of the structure of the universe. Smoot *et al* [27] summarize the different velocities of our Solar system (the Earth) relative to the cosmic blackbody radiation, nearby galaxies and the Milky Way galaxy; also the motion of the Milky Way galaxy relative to the cosmic blackbody radiation. Therefore, the Earth experiences a significant motion relative to the rest frame, which is termed the absolute velocity $\mathbf{v}$ of the Earth.

Velocity addition is a consequence of the Galilean transformations (Newtonian Mechanics) which is also everyday experience that builds our "common sense". Let a moving frame $S(txyz)$ be attached to the Earth which has a velocity **v** relative to the ether frame $\Sigma(TXYZ)$. If **c** is the velocity of light in the ether frame $\Sigma(TXYZ)$, then according to Galilean transformation the velocity of light in the moving frame $S(txyz)$ (the Earth frame) is

$$\mathbf{c}(\theta, v) = \mathbf{c} + \mathbf{v} \qquad \text{(Velocity addition)} \tag{1}$$

where the magnitude of the speed of light $c \approx 3 \times 10^8$ $\text{ms}^{-1}$ (in vacuum).

Following Lämmerzahl [28] we can write the orientation and velocity dependent modulus:

$$\begin{aligned} c(\theta, v) &= [c^2 + v^2 + 2cv\cos\theta]^{\frac{1}{2}} \\ &\approx c\left[1 + \frac{v}{c}\cos\theta + \frac{1}{2}\frac{v^2}{c^2}\sin^2\theta\right] \end{aligned} \tag{2}$$

after omitting 3rd and higher order terms of $\left(\frac{v}{c}\right)$ and where $\theta$ is the angle between the propagation direction of light **c** and the absolute velocity of the Earth (or laboratory) **v**. Therefore, a violation of the constancy of the speed of light would imply an orientation dependence and velocity dependence of $c(\theta, v)$. The Michelson-Morley [3] experiment sought to identify the orientation dependence and the Kennedy-Thorndike [17] experiment examined the velocity dependence of $c(\theta, v)$. This ether drift theoretical framework was used by Michelson and Morley at the time of their investigation in 1887.

## 2.2. The Cosmic Microwave Background (CMB) frame :

According to modern cosmological theory, the initial starting point of the Universe was a Big Bang from which the Universe expanded from a very hot, dense phase about 15 billion years ago. The radiation from this point in time has now cooled to a blackbody temperature of $2.73^0$ K and is identified as the Cosmic Microwave Background (CMB). In 1965, Penzias and Wilson [26] were the first to detect this CMB and also reported an isotropic character of CMB.

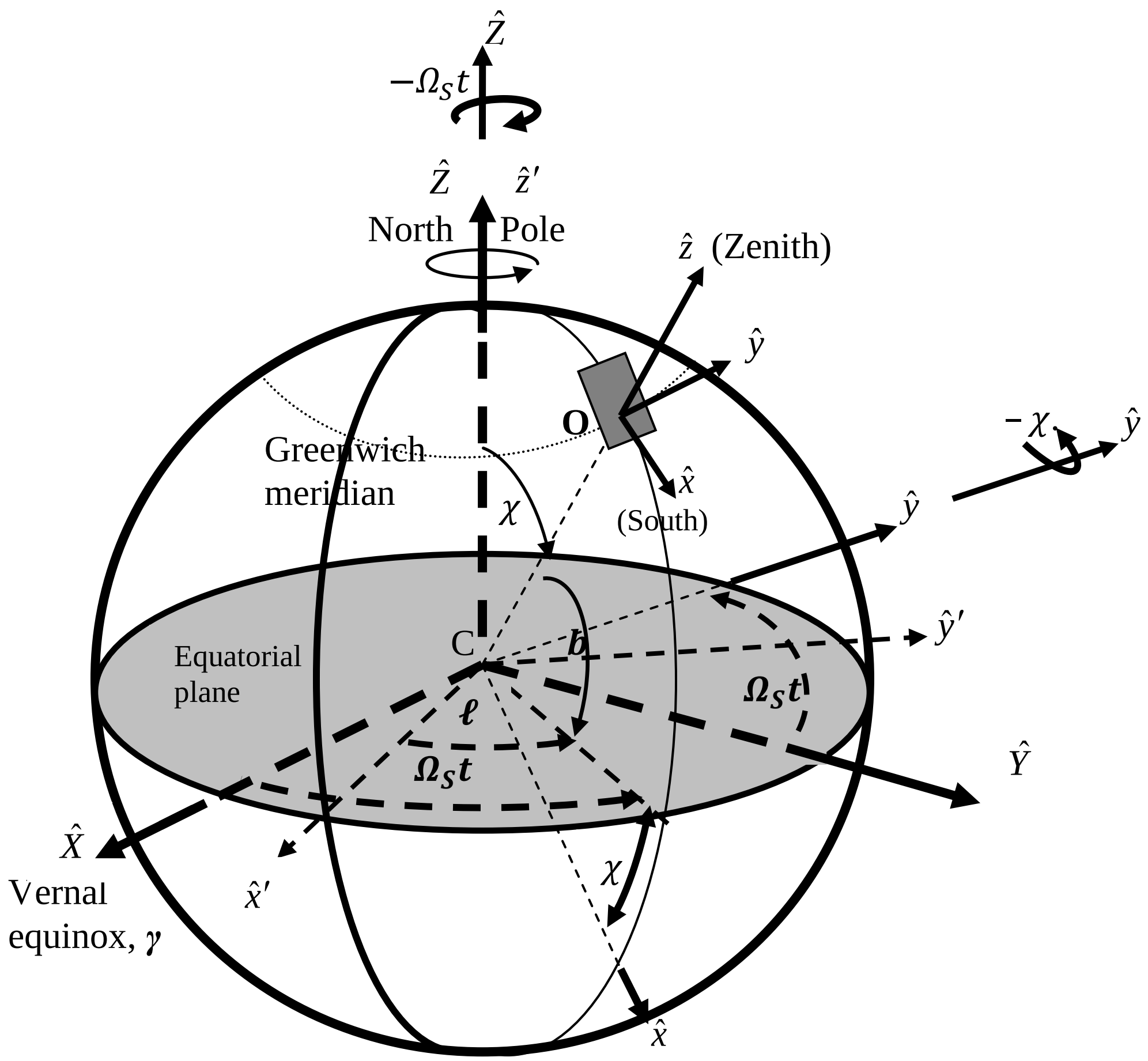

Figure 1. Schematic diagram of the Earth-centered inertial frame $\mathrm{C}(XYZ)$; the Earth-centered non-inertial frame $\mathrm{C}(x'y'z')$ embedded in and rotating with the Earth; and any arbitrary Earth based laboratory frame $\mathrm{O}(xyz)$ with (longitude, latitude) = $(\ell, b)$ and co-latitude, $\chi = \frac{\pi}{2} - b$ centered at C of the Earth's center and rotating with the Earth's axis with sidereal angular rotational velocity $\Omega_S$. The time $t = 0$ starts on the first day of autumn 21st September (Autumnal Equinox). In order to derive the rotation matrices to make the transformation between $\mathrm{C}(XYZ)$ and $\mathrm{O}(xyz)$ frames, the rotation angles $-\chi$ and $-\Omega_\mathrm{S} t$ with rotation axis $\hat{y}$ and $\hat{Z}$ respectively have been shown.

The preferred frame of reference is identified with the CMB frame in the modern version of experiments to test the isotropy of the speed of light. Therefore, the moving frame of reference attached with the Earth which represents our laboratory system is moving with a velocity $\geq 3 \times 10^5\ \mathrm{ms}^{-1}$ relative to CMB [27, 29 - 33]. This would lead to an improved limit of the verification of the isotropy of the speed of light. The CMB reference frame may be described as follows:

Let the Earth-centered inertial frame $\mathrm{C}(XYZ)$ be defined as: the $X-$axis along the vernal equinox (0° Right Ascension (RA) and 0° Declination (Dec)), the $Z-$axis pointing towards the Earth's North Pole, the Earth's axis of rotation (90° Dec), and the $Y-$axis is at $90°$ RA and $0°$ Dec, taken in the J2000.0 frame [34]. Also any arbitrary Earth based laboratory coordinates $\mathrm{O}(xyz)$ are defined at the point of the experiment where ***x***-axis points south, ***y***-axis points east and the ***z***-axis points towards the zenith as shown in Fig. 1. The laboratory frame $\mathrm{O}(xyz)$ rotates with the Earth's axis at a sidereal angular rotation speed $\Omega_\mathrm{S}$. The $\mathrm{C}(xyz)$ frame represents the parallel axis of $\mathrm{O}(xyz)$ frame at the center of the Earth.

Following Fig.1, the rotation matrix $[R_Z(-\Omega_\mathrm{S}t)] = \begin{bmatrix} \cos\Omega_\mathrm{S}t & -\sin\Omega_S\, t & 0 \\ \sin\Omega_\mathrm{S}t & \cos\Omega_\mathrm{S}t & 0 \\ 0 & 0 & 1 \end{bmatrix}$ can be used to make a rotation about the $Z$- axis of the Earth-centered inertial frame $\mathrm{C}(XYZ)$ through an angle $-\Omega_\mathrm{S}t$ and the rotation matrix $\left[R_y(-\chi)\right] = \begin{bmatrix} \cos\chi & 0 & \sin\chi \\ 0 & 1 & 0 \\ -\sin\chi & 0 & \cos\chi \end{bmatrix}$ can be used to make a rotation about the $y$-axis of $\mathrm{C}(xyz)$ at the center of the Earth through an angle $-\chi$.

Using the rotation matrices $[R_Z(-\Omega_\mathrm{S}t)]$ and $\left[R_y(-\chi)\right]$, we can derive the transformation matrix from the $\mathrm{O}(xyz)$ frame into the $\mathrm{C}(XYZ)$ frame as

$$[T]_{xX} = [R_Z(-\Omega_S t)]\left[R_y(-\chi)\right] = \begin{bmatrix} \cos\chi\cos\Omega_S t & -\sin\Omega_S t & \sin\chi\cos\Omega_S t \\ \cos\chi\sin\Omega_S t & \cos\Omega_S t & \sin\chi\sin\Omega_S t \\ -\sin\chi & 0 & \cos\chi \end{bmatrix} \tag{3}$$

This is an orthogonal matrix, so that for the inverse transformation from the $\mathrm{C}(XYZ)$ frame into the $\mathrm{O}(xyz)$ frame we use $[T]_{Xx} = ([T]_{xX})^T$ [34] or

$$[T]_{Xx} = \begin{bmatrix} \cos\chi\cos\Omega_S t & \cos\chi\sin\Omega_S t & -\sin\chi \\ -\sin\Omega_S t & \cos\Omega_S t & 0 \\ \sin\chi\cos\Omega_S t & \sin\chi\sin\Omega_S t & \cos\chi \end{bmatrix} \tag{4}$$

The Sun -Centered Celestial Equatorial Frame (SCCEF) has been elaborated in detail in [20, 35 - 40] attempts to explain tests of the Lorentz Invariance. The SCCEF$(XYZ)$ is the frame in which the Sun is at the centre, and is inertial relative to the CMB$(X_\Sigma Y_\Sigma Z_\Sigma)$ frame to first order. The axis in the SCCEF$(XYZ)$ are defined as shown in Fig. 2: the $Z$-axis is parallel to the Earth's

North Pole (90° Dec), the $X$-axis is at 0° RA and 0° Dec which points from the Sun toward the Earth at the moment of the autumnal equinox, while the Y-axis is at 90° RA and 0° Dec, taken in the J2000.0 frame. The axis in the CMB($X_\Sigma Y_\Sigma Z_\Sigma$) frame are defined as shown in Fig.2: the $\hat{X}_\Sigma$ pointing towards $(\alpha, \delta) = (168°, -7.22°)$, $\hat{Z}_\Sigma$ pointing towards $(\alpha, \delta') = (168°, 90° - 7.22°)$ and the $\hat{Y}_\Sigma$ - axis completes the right-handed system where $\alpha$ = RA and $\delta$ = Dec.

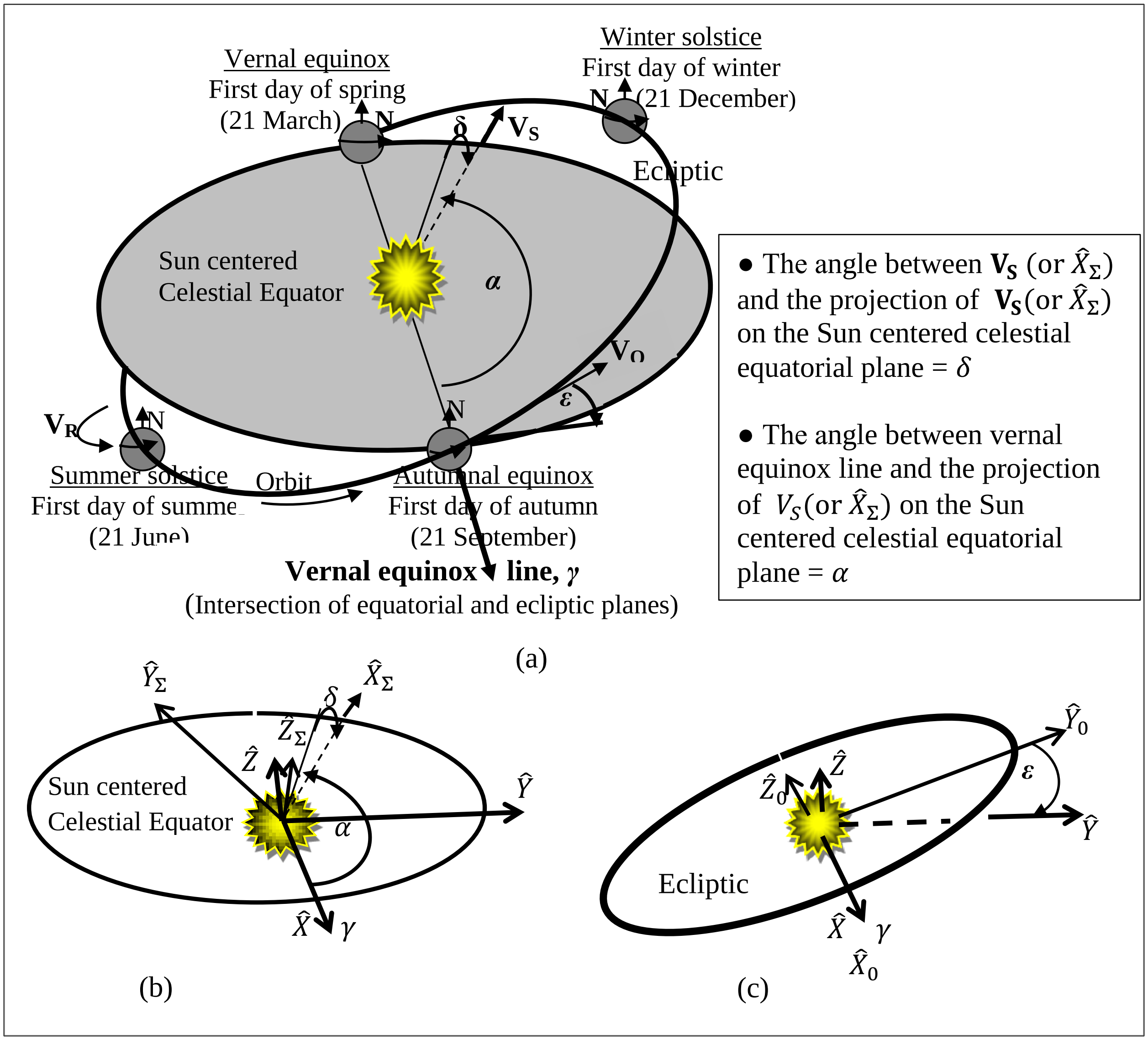

Figure 2. Schematic diagram of (a) Movements of the Earth in space showing different velocities described in the Table-1, viewed from above the celestial equatorial plane (which also show the change of seasons in the northern hemisphere) [34]. The angle between the Ecliptic (orbital plane) and the Sun centered Celestial Equatorial plane is $\varepsilon \approx 23.4°$. (b)The Cosmic Microwave Background (CMB) $(X_\Sigma Y_\Sigma Z_\Sigma)$ -frame and the Sun-Centered Celestial Equatorial Frame (SCCEF)$(XYZ)$. (c) The barycentric non-rotating frame (BRS)$(X_O Y_O Z_O)$ and the Sun-Centered Celestial Equatorial Frame (SCCEF)$(XYZ)$.

Fig. 2(b) represents a schematic diagram of the SCCEF$(XYZ)$ frame and the CMB$(X_\Sigma Y_\Sigma Z_\Sigma)$ frame. The Earth-centered inertial frame $C(XYZ)$ presented in Fig.1 has the same axis orientation as the SCCEF$(XYZ)$ frame in Fig.2. Therefore, the SCCEF$(XYZ)$ frame and the Earth-centered inertial frame $C(XYZ)$ are parallel frames.

Following Fig. 2(b) the rotation matrix $\left[R_{Y_\Sigma}(-\delta)\right] = \begin{bmatrix} \cos\delta & 0 & \sin\delta \\ 0 & 1 & 0 \\ -\sin\delta & 0 & \cos\delta \end{bmatrix}$ can be used to make the rotation about the $Y_\Sigma$- axis of the CMB$(X_\Sigma Y_\Sigma Z_\Sigma)$ frame with an angle $(-\delta)$ and the rotation matrix $[R_Z(-\alpha)] = \begin{bmatrix} \cos\alpha & -\sin\alpha & 0 \\ \sin\alpha & \cos\alpha & 0 \\ 0 & 0 & 1 \end{bmatrix}$ can be used to make the rotation about the $Z$-axis of the SCCEF$(XYZ)$ frame with an angle $(-\alpha)$.

Using the rotation matrices $\left[R_{Y_\Sigma}(-\delta)\right]$ and $[R_Z(-\alpha)]$, we can derive the transformation matrix from the CMB$(X_\Sigma Y_\Sigma Z_\Sigma)$ into the SCCEF$(XYZ)$ frame as

$$[T]_{X_\Sigma X} = [R_Z(-\alpha)]\left[R_{Y_\Sigma}(-\delta)\right] = \begin{bmatrix} \cos\alpha\cos\delta & -\sin\alpha & \cos\alpha\sin\delta \\ \sin\alpha\cos\delta & \cos\alpha & \sin\alpha\sin\delta \\ -\sin\delta & 0 & \cos\delta \end{bmatrix} \tag{5}$$

Also in order to calculate the orbital velocity $\mathbf{V_O}$, we consider the earth in a barycentric non-rotating frame (BRS)$(X_O Y_O Z_O)$ [35] where the spatial origin coincides with the centre of the Sun with the $\hat{Z}_O$- axis perpendicular to the Ecliptic (Orbital plane), the $\hat{X}_O$- axis points to the vernal equinox line, and the $\hat{Y}_O$- axis completes the right-handed system. Fig. 2(c) represents a schematic diagram of the BRS$(X_O Y_O Z_O)$ and the SCCEF$(XYZ)$ frames.

### 2.2.1. The time dependent expressions of the velocity of the laboratory :

In order to derive the time dependent expressions of the velocity of the laboratory $\mathbf{V}(t)$, all contributing velocities $(\mathbf{V_R}(t), \mathbf{V_O}(t), \mathbf{V_S}(t))$ which have been described in Table-1, Fig.1 and Fig. 2, are transformed into the inertial SCCEF$(XYZ)$ frame. Galilean transformations for the velocities are sufficient for these calculations as the velocities are much smaller than the speed of light which have been presented in the Table-1. In order to calculate the orbital velocity $\mathbf{V}_O$, we first consider the Earth in a barycentric non-rotating frame (BRS)$(X_O Y_O Z_O)$ [35] which has been described in the previous section in Fig. 2(c).

The Earth's orbital velocity in the BRS$(X_O Y_O Z_O)$ frame can be derived as:

$$\mathbf{V}_{\mathrm{O_BRS}} = V_O \begin{bmatrix} \cos(\pi/2 + \Omega_O t) \\ \sin(\pi/2 + \Omega_O t) \\ 0 \end{bmatrix} = V_O \begin{bmatrix} -\sin\Omega_O t \\ \cos\Omega_O t \\ 0 \end{bmatrix} \tag{6}$$

In order to transform the Earth's orbital velocity $\mathbf{V}_{O_BRS}$ in the BRS$(X_O Y_O Z_O)$ frame into the SCCEF$(XYZ)$ frame we will use an orthogonal transformation matrix associated with the rotation about the common $X$-axis with the angle $\varepsilon$ as shown in the Fig. 2:

$$[R_X(-\varepsilon)] = \begin{bmatrix} 1 & 0 & 0 \\ 0 & \cos\varepsilon & -\sin\varepsilon \\ 0 & \sin\varepsilon & \cos\varepsilon \end{bmatrix} \tag{7}$$

Using (6) and (7), we can derive the Earth's orbital velocity in the SCCEF$(XYZ)$ frame as:

$$\mathbf{V}_{\mathrm{O_SCCEF}} = \begin{bmatrix} 1 & 0 & 0 \\ 0 & \cos\varepsilon & -\sin\varepsilon \\ 0 & \sin\varepsilon & \cos\varepsilon \end{bmatrix} \begin{bmatrix} -V_O \sin\Omega_O t \\ V_O \cos\Omega_O t \\ 0 \end{bmatrix} = \begin{bmatrix} -V_O \sin\Omega_O t \\ V_O \cos\varepsilon \cos\Omega_O t \\ V_O \sin\varepsilon \cos\Omega_O t \end{bmatrix} \tag{8}$$

We can derive the Earth's rotational velocity in the laboratory-frame O$(xyz)$ which was described in the Fig.1 as:

$$\mathbf{V}_{\mathrm{R_Lab}} = V_R \begin{bmatrix} 0 \\ 1 \\ 0 \end{bmatrix} \tag{9}$$

The transformation matrix $|T|_{xX}$ has been derived to transform the laboratory-frame O$(xyz)$ into the SCCEF$(XYZ)$ frame in (3). Using (3) and (9), we derive the rotational velocity of the earth in the SCCEF$(XYZ)$ as:

$$\mathbf{V}_{\mathrm{R_SCCEF}} = \begin{bmatrix} \cos\chi\cos\Omega_S t & -\sin\Omega_S t & \sin\chi\cos\Omega_S t \\ \cos\chi\sin\Omega_S t & \cos\Omega_S t & \sin\chi\sin\Omega_S t \\ -\sin\chi & 0 & \cos\chi \end{bmatrix} \begin{bmatrix} 0 \\ V_R \\ 0 \end{bmatrix} = \begin{bmatrix} -V_R\sin\Omega_S t \\ V_R\cos\Omega_S t \\ 0 \end{bmatrix} \tag{10}$$

We can derive the velocity of the solar system relative to the CMB($X_\Sigma Y_\Sigma Z_\Sigma$) frame as:

$$\mathbf{V}_{S_CMB} = V_S \begin{bmatrix} 1 \\ 0 \\ 0 \end{bmatrix} \tag{11}$$

The transformation matrix $[T]_{X_\Sigma X}$ has been derived to transform the CMB($X_\Sigma Y_\Sigma Z_\Sigma$) frame into the SCCEF($XYZ$) frame in (5). Using (5) and (11), we derive the velocity of the solar system in the SCCEF($XYZ$) as:

$$\mathbf{V}_{S_SCCEF} = \begin{bmatrix} \cos\alpha\cos\delta & -\sin\alpha & \cos\alpha\sin\delta \\ \sin\alpha\cos\delta & \cos\alpha & \sin\alpha\sin\delta \\ -\sin\delta & 0 & \cos\delta \end{bmatrix} \begin{bmatrix} V_S \\ 0 \\ 0 \end{bmatrix} = \begin{bmatrix} V_S\cos\alpha\cos\delta \\ V_S\sin\alpha\cos\delta \\ -V_S\sin\delta \end{bmatrix} \tag{12}$$

In order to derive total velocity of laboratory in the SCCEF($XYZ$) frame, we will add equations (8), (10) and (12) as follows:

$$\mathbf{V}_{SCCEF}(t) = \begin{bmatrix} V_{SCCEF_X} \\ V_{SCCEF_Y} \\ V_{SCCEF_Z} \end{bmatrix} = \begin{bmatrix} -V_O\sin\Omega_O t - V_R\sin\Omega_S t + V_S\cos\alpha\cos\delta \\ V_O\cos\varepsilon\cos\Omega_O t + V_R\cos\Omega_S t + V_S\sin\alpha\cos\delta \\ V_O\sin\varepsilon\cos\Omega_O t - V_S\sin\delta \end{bmatrix} \tag{13}$$

### 2.2.2. The time dependent expressions of the angle between the direction of the light propagation and the direction of the velocity of the laboratory :

According to the direction of the propagation of light [along North-South (N-S), or East-West (E-W) or (Zenith) (Z)] in the laboratory-frame O($xyx$), we can derive the unit vectors as follows [33]:

$$\hat{e}_{lab,N-S} = \begin{bmatrix} 1 \\ 0 \\ 0 \end{bmatrix}; \qquad \hat{e}_{lab,E-W} = \begin{bmatrix} 0 \\ 1 \\ 0 \end{bmatrix}; \qquad \hat{e}_{lab,Zenith} = \begin{bmatrix} 0 \\ 0 \\ 1 \end{bmatrix} \tag{14}$$

The transformation from the laboratory-frame O$(xyz)$ into the SCCEF$(XYZ)$ is performed using the transformation matrix $[T]_{xX}$ in equation (3) as:

$$\hat{e}_1(t) = \begin{bmatrix} \cos\chi \cos\Omega_S t \\ \cos\chi \sin\Omega_S t \\ -\sin\chi \end{bmatrix}; \text{ North-South is the propagation direction of light} \quad (15a)$$

$$\hat{e}_2(t) = \begin{bmatrix} -\sin\Omega_S t \\ \cos\Omega_S t \\ 0 \end{bmatrix}; \text{ East-West is the propagation direction of light} \quad (15b)$$

$$\hat{e}_3(t) = \begin{bmatrix} \sin\chi \cos\Omega_S t \\ \sin\chi \sin\Omega_S t \\ \cos\chi \end{bmatrix}; \text{ Zenith is the propagation direction of light} \quad (15c)$$

Using (13) we can derive the unit vector of the velocity of the laboratory relative to the CMB$(X_\Sigma Y_\Sigma Z_\Sigma)$ frame as:

$$\hat{V}_{SCCEF}(t) = \frac{\mathbf{V}_{SCCEF}}{|\mathbf{V}_{SCCEF}|} = \frac{1}{|\mathbf{V}_{SCCEF}|} \begin{bmatrix} -V_O \sin\Omega_O t - V_R \sin\Omega_S t + V_S \cos\alpha \cos\delta \\ V_O \cos\varepsilon \cos\Omega_O t + V_R \cos\Omega_S t + V_S \sin\alpha \cos\delta \\ V_O \sin\varepsilon \cos\Omega_O t - V_S \sin\delta \end{bmatrix} \quad (16)$$

Using (13) and (15a-15c), we can derive the components of the velocity of the laboratory relative to CMB along the direction of the propagation of light in the SCCEF$(XYZ)$ frame as shown in the Table-3.

The graphical presentation of the equations in Table-3 gives the predicted hypothetical variation of the speed of light with time over the year – including the first order terms in $(v/c)$ and also, after some derivation, the second order terms in $(v/c)$. For example, if we perform any isotropy experiment at York University, Toronto, Ontario, Canada then using the values of the parameters presented in the Table-1 and Table-2, we can produce the graphical presentation of the time dependent components of the velocity of the laboratory relative to the CMB along the direction of the light propagation in the SCCEF$(XYZ)$ frame as shown in Figs. 3 and 4.

Table 1. Movements of the Earth compared to the speed of light, $\beta_i = \frac{v_i}{c}$ where $v_i$ is the different velocities (rotational, orbital and Sun's) of the Earth and $c (c \approx 3 \times 10^8 \; ms^{-1})$ is the speed of light in vacuum.

| | |
|---|---|
| Earth's spin motion: | The boost speed of the laboratory on the Earth's surface due to its spin motion is $\beta_R = \frac{V_R}{c} = (0 \leq \beta_R \leq 1.5 \times 10^{-6})$; where $V_R (0 \leq V_R \leq 4.5 \times 10^2 \text{ ms}^{-1})$ is the velocity due to the Earth's rotation about its axis depending on the geographical latitude. The Earth is rotating relative to its axis with sidereal angular rotational frequency $\Omega_S = \frac{2\pi}{23 \text{ h } 56 \text{ min}} \cong 4.18 \times 10^{-3} \text{degree/sec}$ [35, 36]. |
| Earth's orbital motion: | The boost speed of the laboratory on the Earth's surface due to the Earth's orbital motion is $\beta_O = \frac{V_O}{c} \approx 10^{-4}$; where $V_O = 3 \times 10^4 ms^{-1}$ is the velocity due to the Earth's orbital motion relative to the Sun. The Earth is orbiting relative to the Sun with the angular frequency $\Omega_O = \frac{2\pi}{1 \text{ yr}} \approx 1.14 \times 10^{-5} \text{degree/sec}$ [35, 36]. |
| Sun's motion relative to CMB: | The boost speed of the laboratory on the Earth's surface due to the velocity of the solar system relative to the CMB is $\beta_S = \frac{V_S}{c} \approx 10^{-3}$; where $V_S \approx 3.71 \times 10^5 \; ms^{-1}$ is the velocity of the solar system towards $[(\alpha, \delta) = (168°, -7.22°)]$ relative to the CMB, where $\alpha =$ right ascension and $\delta =$ declination [29 - 33]. |

Table 2. The geographical dependent parameters and its numerical values for Toronto, Canada.

| | |
|---|---|
| Spin motion: | $V_R \cong 2.4 \times 10^2 \text{ ms}^{-1}$ |
| Longitude: | $\ell \cong 79°24'$ West |
| Latitude: | $b \cong 43°40'$ North |
| Co-latitude: | $\chi = \frac{\pi}{2} - b \cong 46°20'$ |

Table 3. The time dependent components of the velocity of the laboratory relative to the CMB along the direction of the light propagation in the SCCEF($XYZ$) frame.

| Propagation direction of light in laboratory | The components of the velocity of the laboratory relative to CMB along the direction of the light propagation = $\lvert\mathbf{V}_{SCCEF}\rvert \cos\theta_i$ where $\theta_i$ is the angle between the propagation direction of light and the absolute velocity of the Earth (or laboratory). |
|---|---|
| North-South | $\{\cos\chi\cos\Omega_S t\}\{-V_O\sin\Omega_O t - V_R\sin\Omega_S t + V_S\cos\alpha\cos\delta\} + \{\cos\chi\sin\Omega_S t\}\{V_O\cos\varepsilon\cos\Omega_O t + V_R\cos\Omega_S t + V_S\sin\alpha\cos\delta\} + \{-\sin\chi\}\{V_O\sin\varepsilon\cos\Omega_O t - V_S\sin\delta\}$ |
| East-West | $\{-\sin\Omega_S t\}\{-V_O\sin\Omega_O t - V_R\sin\Omega_S t + V_S\cos\alpha\cos\delta\} + \{\cos\Omega_S t\}\{V_O\cos\varepsilon\cos\Omega_O t + V_R\cos\Omega_S t + V_S\sin\alpha\cos\delta\}$ |
| Zenith | $\{\sin\chi\cos\Omega_S t\}\{-V_O\sin\Omega_O t - V_R\sin\Omega_S t + V_S\cos\alpha\cos\delta\} + \{\sin\chi\sin\Omega_S t\}\{V_O\cos\varepsilon\cos\Omega_O t + V_R\cos\Omega_S t + V_S\sin\alpha\cos\delta\} + \{\cos\chi\}\{V_O\sin\varepsilon\cos\Omega_O t - V_S\sin\delta\}$ |

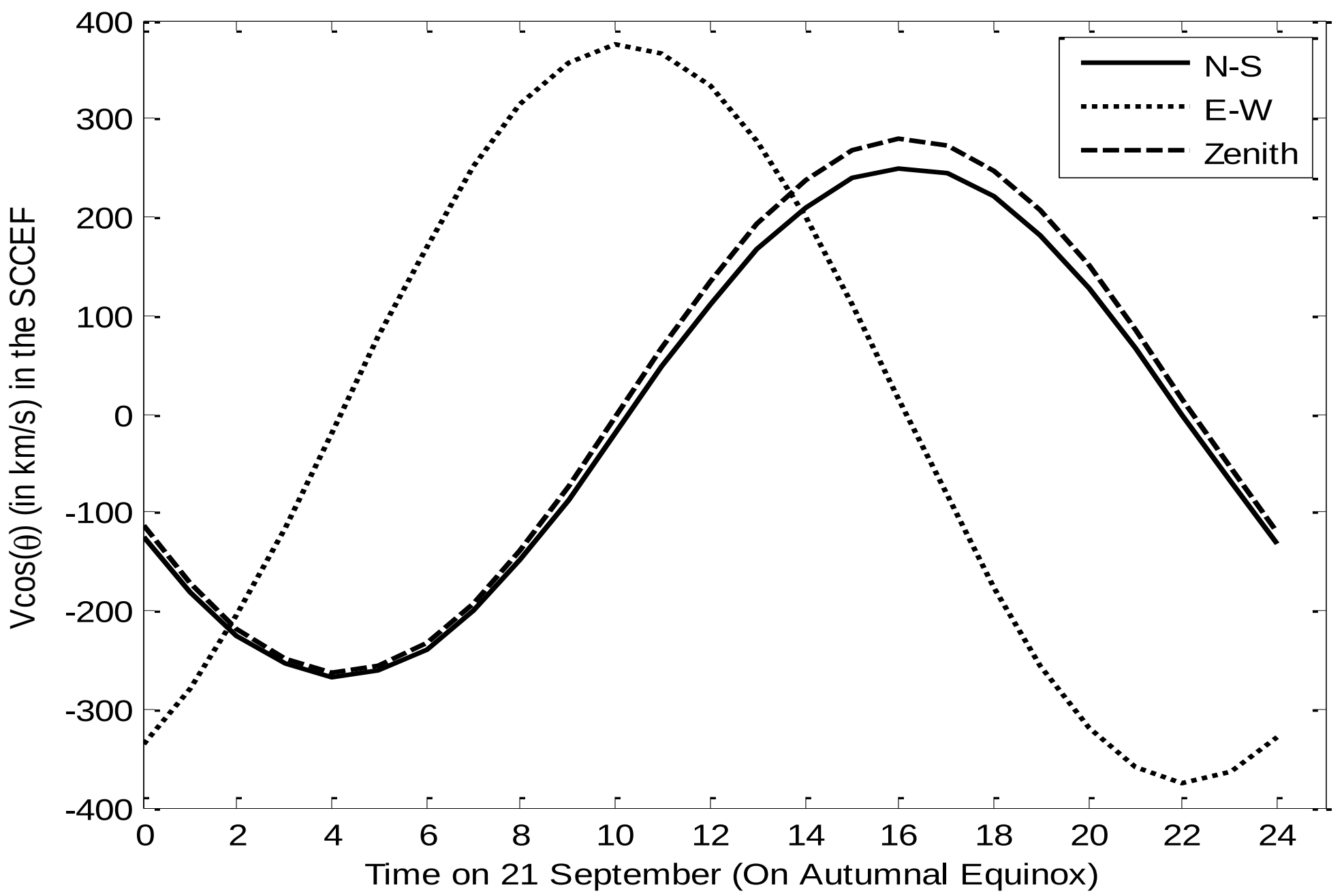

Figure 3.The component of the velocity of the laboratory $\lvert\mathbf{V}_{SCCEF}\rvert\cos\theta$ in the direction of the light propagation at different times on 21$^{st}$ September, where $\theta$ is the angle between the direction of the velocity of the laboratory and the direction of the propagation of light.

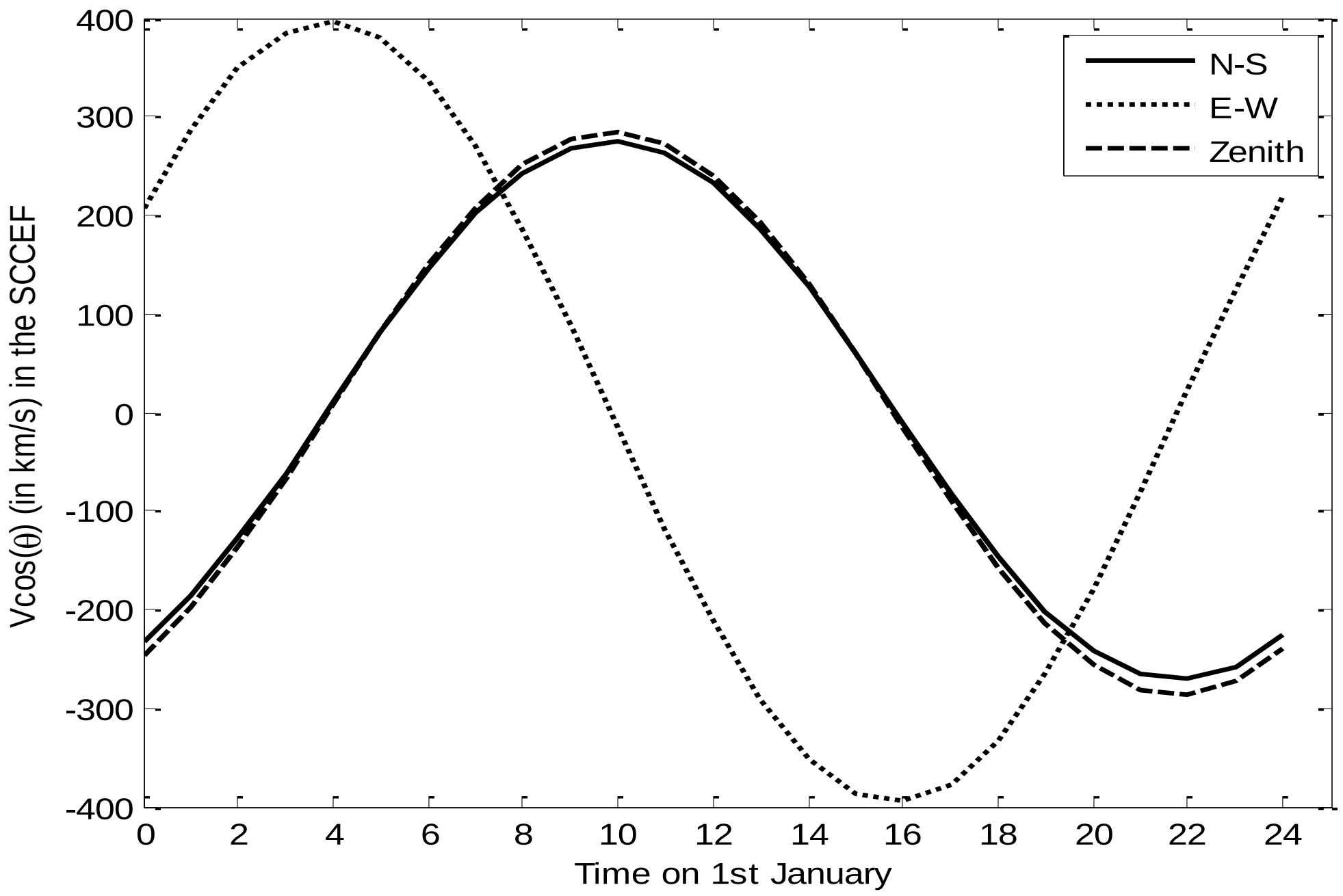

Figure 4. The component of the velocity of the laboratory $|\mathbf{V}_{SCCEF}|\cos\theta$ in the direction of the light propagation at different times on 1st January, where $\theta$ is the angle between the direction of the velocity of the laboratory and the direction of the propagation of light.

## 2.3. Kinematical frameworks to interpret experimental tests :

In order to parameterize and identify the possible violations of Special Relativity while quantifying the degree of validity, different frameworks (test theories) have been proposed by different authors. Recently these test theories have been discussed further in [22, 28]. These theories are an important step for the understanding of the structure of the special theory of relativity as well as being very useful to compare the results of different experiments to test the validity of special theory of relativity. For the purpose of quantifying different experiments to test the isotropy of the speed of light which have been reviewed in this article, we adopt the widely used kinematical test theories of Robertson [13], and Mansouri and Sexl [14 - 16]. These are generally called the RMS-test theories [28, 41, 42] and are discussed further in [19, 43]. Presently another widely used dynamical test theory to describe experiments is the Standard Model Extension (known as SME) [44]. However, for our present purpose we will concentrate on the RMS-framework.

### 2.3.1. Robertson-Mansouri-Sexl (RMS)-test theories :

In order to give an idea of these test theories compared with Galilean (Newtonian) and Lorentz transformations, let us consider two inertial reference frames $\Sigma(TXYZ)$ and $S(txyz)$ where $\Sigma$ is the hypothetical rest frame. Therefore the speed of light is isotropic in this hypothetical rest frame $\Sigma$. The $S$-frame is moving at a uniform velocity **v** along X-axis relative to the $\Sigma$-frame. In order to transfer the time and the Cartesian coordinates of a physical phenomenon from the $S(txyz)$ - frame into the $\Sigma(TXYZ)$ -frame choosing $T = t = 0$ as the common origin, we can write as follows:

$$X^{\mu} = \sum_{i=0}^{3} \eta_i^{\mu}(v) x^i \tag{17}$$

where $(X^{\mu}) = (T; X, Y, Z)$ and $(x^i) = (t; x, y, z)$ consist of temporal coordinates $(X^0 = T; x^0 = t)$ and spatial Cartesian coordinates $(X^1 = X; X^2 = Y; X^3 = Z$ and $x^1 = x; x^2 = y; x^3 = z)$.

Following [13, 14, 20, 22, 28], the general transformations between $\Sigma(TXYZ)$ and $S(txyz)$ -frames with arbitrary synchronization of clocks are as follows:

$$\begin{aligned} T &= \frac{\gamma}{a(v^2)}(t - \boldsymbol{\epsilon}.\mathbf{x}) \\ X &= \frac{1}{d(v^2)}x - \left(\frac{1}{d(v^2)} - \frac{\gamma}{b(v^2)}\right)\frac{v(\mathbf{v}.\mathbf{x})}{v^2} - \frac{1}{a(v^2)}v(\boldsymbol{\epsilon}.\mathbf{x}) + \frac{\gamma}{a(v^2)}vt \\ Y &= \frac{y}{d(v^2)}; \quad Z = \frac{z}{d(v^2)} \end{aligned} \tag{18}$$

where $a(v^2), b(v^2)$ and $d(v^2)$ are test functions and $\boldsymbol{\epsilon}$ is a vector determined by the procedure adopted for the global clocks synchronizations in the *S*-frame. Also, $\gamma = 1/\sqrt{1-(v/c)^2}$.

Since $\Sigma(TXYZ)$ is a hypothetical preferred inertial reference frame (CMB is the best candidate), in which the speed of light is isotropic, therefore

$$c^2T^2 - X^2 - Y^2 - Z^2 = 0 \tag{19}$$

Setting equation (19) in the $\Sigma(TXYZ)$ frame and using the general form of the Lorentz transformations with arbitrary synchronization of clocks shown in equation (18), we can derive the speed of light in $S(txyz)$ frame as $c = c(v, \theta, \epsilon)$ where $\theta$ is the angle between the propagation direction of light and the absolute velocity of the Earth (or laboratory). Therefore according to the generalized test theory with arbitrary clocks synchronizations the velocity of light not only depends on the orientation and velocity of the source but also the synchronization condition used. A brief review of the history of clocks synchronizations are discussed in [14].

#### 2.3.1.1. The round-trip (two-way) velocity of light :

Using the RMS-test theory as described in Table-4 and following [45 - 47], we can parameterize the orientation and velocity dependence of the two-way speed of light as follows:

$$
\begin{aligned}
c(v,\theta) &= c\left[1 + (\alpha - \beta + 1)\frac{v^2}{c^2} + \left(\beta - \delta - \frac{1}{2}\right)\frac{v^2}{c^2}\sin^2\theta + \mathcal{O}(c^{-4})\right] \\
&\cong c\left[1 + \left\{(\alpha - \beta + 1) + \left(\beta - \delta - \tfrac{1}{2}\right)\sin^2\theta\right\}\tfrac{v^2}{c^2}\right] \qquad (20)
\end{aligned}
$$

where $c$ is the constant speed of light in the $\Sigma$ −frame and $\theta$ is the angle between the direction of light propagation and the velocity vector of the $S$-frame relative to $\Sigma$ −frame. If the STR is valid then, $(\alpha - \beta + 1) = \left(\beta - \delta - \frac{1}{2}\right) = 0$, i.e. $\left[\alpha = -\frac{1}{2};\ \beta = \frac{1}{2}, \text{and } \delta = 0\right]$. A Michelson-Morley type experiment can set upper limits on $[(\beta - \delta - \frac{1}{2})]$; a Kennedy-Thorndike type experiment can set upper limits on $[1 + (\alpha - \beta + 1) + (\beta - \delta - \frac{1}{2})\sin^2\theta]$.

Equation (20) represents the two-way velocity of light as it is derived under the assumption of Einstein synchronization where the round trip speed of light has been considered. According to general test theory [14, 28], the one-way velocity of light depends on the synchronization parameter. However, Will [49] showed that experiments which test the isotropy in one-way or two-way (round-trip) experiments have observables that depend on test functions $a(v^2), b(v^2)$ and $d(v^2)$ but not on the synchronization procedure. He noted that “the synchronization of clocks played no role in the interpretation of experiments provided that one is careful to express the results in terms of physically measurable quantities”. Hence the synchronization is largely irrelevant. We will use RMS-test theory for this review paper to describe the experiments.

Table 4. Different assumptions and consequences for widely used frame transformations. **[$\gamma = (1 - v^2/c^2)^{-1/2}$; RMS**= Robertson, and Mansouri and Sexl [13 - 16]].

| **Frames transformations [Equation 17]** | **Consequences /Assumptions** | **Clocks Synchronization ( use absolute simultaneity,** $T = t = 0$) | **Comments** |
|---|---|---|---|
| **Galilean**<br>$T = t$;<br>$X = x + vt$;<br>$Y = y$; $Z = z$ | (1) No speed limit in nature.<br>(2) $c = c(v, \theta)$. | $T = t =$constant in both frames | (1) Classical Mechanics ($v \ll c$) are invariant.<br>(2) Maxwell's equations are not invariant. |
| **Lorentz**<br>$T = \gamma\left(t + \frac{v}{c^2}x\right)$;<br>$X = \gamma(x + vt)$;<br>$Y = y$; $Z = z$ | (1) There is speed limit in nature.<br>(2) $c =$ const. | No need for round trip speed of light. | (1) Classical Mechanics ($v \ll c$) are invariant.<br>(2) Maxwell's equations are invariant. |
| **Robertson**<br>$T = a_0^0 t + \frac{v}{c^2} a_1^1 x$;<br>$X = v a_0^0 t + a_1^1 x$;<br>$Y = a_2^2 y$; $Z = a_2^2 z$ | (1) $c$ is isotropic only for hypothetical preferred Σ-frame.<br>(2) $c = c(v, \theta)$ need to test experimentally. | Einstein synchronization [13, 14] | $g_0 = [1 - (v/c)^2]^{1/2} a_0^0 = \frac{a_0^0}{\gamma}$;<br>$g_1 = [1 - (v/c)^2]^{1/2} a_1^1 = \frac{a_1^1}{\gamma}$;<br>$g_2 = a_2^2$;<br>where, for STR<br>$g_0 = g_1 = g_2 = 1$ |
| **RMS**[1]<br>$T = \gamma\left(\frac{t}{a(v^2)} + \frac{vx}{b(v^2)c^2}\right)$<br>$X = \gamma\left(\frac{x}{b(v^2)} + \frac{vt}{a(v^2)}\right)$<br>$Y = \frac{y}{d(v^2)}$; $Z = \frac{z}{d(v^2)}$ | (1) $c$ is isotropic only for hypothetical preferred Σ-frame.<br>(2) $c = c(v, \theta)$ need to test experimentally. | Einstein synchronization [13,14] | Test functions are derived in the low-velocity limit as [4,8]<br>$a(v^2) = [1 + \alpha\left(\frac{v}{c}\right)^2 + \mathcal{O}(c^{-4})]$<br>$b(v^2) = [1 + \beta\left(\frac{v}{c}\right)^2 + \mathcal{O}(c^{-4})]$<br>$d(v^2) = [1 + \delta\left(\frac{v}{c}\right)^2 + \mathcal{O}(c^{-4})]$<br>and also;<br>$a(v^2) = \frac{1}{a_0^0} = \frac{1}{\gamma g_0}$<br>$b(v^2) = \frac{\gamma^2}{a_1^1} = \frac{\gamma}{g_1}$<br>$d(v^2) = \frac{1}{a_2^2} = \frac{1}{g_2}$<br>where, for special relativity<br>$a(v^2) = b(v^2) = d(v^2) = 1$ |

[1]As $a(v^2), b(v^2)$and $d(v^2)$ test functions are assumed to be independent of the relative direction of motion of the $\Sigma$ −frame and $S$ −frame, therefore there are no odd-order terms in the expressions. These test functions parameterize time dilation as well as Lorentz contraction which are the test parameters for experiments. Three kinematical test parameters $\alpha, \beta$ and $\delta$ can be experimentally determined by the three types of tests such as Michelson-Morley type, Kennedy-Thorndike type and Ives-Stilwell type experiments.

### 2.3.1.2. The single-trip (one-way) velocity of light :

Following equation (6.16) of Mansouri-Sexl [14] and also in [49], we can write the one-way velocity of light as measured in the inertial $S(txyz)$- frame in which the laboratory is at rest as follows:

$$c(v,\theta) = c\left[1-(1+2\alpha)\frac{v}{c}\cos\theta + \mathcal{O}(c^{-3})\right]$$
$$\cong c\left[1-(1+2\alpha)\frac{v}{c}\cos\theta\right] \tag{21}$$

Therefore the one-way velocity of light is a measureable quantity and is direction dependent in general. The test parameter $\alpha$ can be tested by different experiments. For the special theory of relativity $\alpha = -\frac{1}{2}$.

### 2.3.2. Standard-Model Extension (SME)-test theory :

The establishment of the Lorentz Invariance, the foundational symmetry of Einstein's relativity theory, made it possible to unify the classical mechanics and Maxwell's electrodynamics. At the fundamental level, all accepted theoretical descriptions of nature are supported by Lorentz symmetry.

Electromagnetism, the weak nuclear force and the strong nuclear force are three of the four fundamental forces in nature, and are well described by the Standard Model of particle physics at the fundamental level. The standard model does not include gravity. The unification of gravity with the other forces requires a quantum field theory [50]. At present the most promising quantum field theory is string (M) theory which is qualitatively different from the standard model of particle physics that it predicts new physics at the Plank scale [44].

In order to look for the possibility that the new physics involves a violation of Lorentz invariance, the generalized Lorentz violating Standard Model Extension (SME) of particle physics has been developed in recent years [38, 44, 51 - 53]. The most general observer-independent quantum field theoretical framework to investigate the violation of Lorentz invariance is the SME. The general form of a Lorentz violating extension to the Lagrangian of the photonic and matter sectors of the SME have been formulated by different authors [37, 38, 44, 52 - 54]. During recent years, test experiments are being described within the SME [66 - 70].

Recently the RMS-framework has been translated into the SME-framework by redefining the length and time intervals specified by the boosted SME rods and clocks to match numerically those of the boosted RMS rods and clocks by Kostelecký and Mewes [38] as follows:

$$(\mathrm{g}_{\mathrm{RMS}})_{\mu\nu}(a,b,d) = (\mathrm{g}_{\mathrm{SME,eff}})_{\mu\nu}(\{k\})$$

where $(\mathrm{g}_{\mathrm{SME,eff}})_{\mu\nu}$is an effective metric that depend on the coefficients $\{k\}$ and $(\mathrm{g}_{\mathrm{RMS}})_{\mu\nu}$is an effective metric that depends on the RMS-test functions $a(v^2), b(v^2)$and $d(v^2)$. For our present purpose, to review all round-trip and single-trip classical and modern isotropy experiments, we adopt the widely used RMS-test theory in this article.

## 3. The outcome of the isotropy experiments

The isotropy of the speed of light implies a directional invariance property of the speed of light. The experiments to test the isotropy can be divided into two categories, those that measure the speed of light over a return path and are sensitive only to second order terms in $(v/c)$ are called round-trip (two-way) experiments, and those that measure the speed of light over a single path and are sensitive to first order terms in $(v/c)$ are called single-trip (one-way) experiments.

### 3.1. The round-trip (two-way) experiments :

A wide variety of classic and modern isotropy experiments, generally known as Michelson-Morley experiments, have been performed since 1881 to test the violation of special relativity where the measurements record the round-trip averaged speed of light. These experiments look for the hypothetical variation in the speed of light as the laboratory apparatus is rotated in space as shown in Fig.5.

#### 3.1.1. Classical experiments :

Fig. 5(a) shows the basic outline of the classical Michelson-Morley experiment. In this experiment, a beam of monochromatic light of wavelength λ, from a source, is split into two beams in two orthogonal directions. After being reflected at the perpendicular distances, these two reflected beams arrive at a detector where interference fringes are observed. If there is any hypothetical dependence of the velocity of the light due to the velocity of the laboratory then one can observe a fringe-shift $\Delta N$ due to optical path differences. If both arms of the interferometer are equal to the length $L$ and our laboratory is moving along the $X$-axis with a velocity $\mathbf{v}$, then the expected time difference as shown in 'experiment-(a)' in Fig.8 is $\Delta t \cong L\frac{v^2}{c^3}$.So we can write the path difference $= \Delta t \times c = L\frac{v^2}{c^2}$. If the interferometer is turned through $90°$, the direction of the velocity of the laboratory, v is unchanged but the two paths of the interferometer will be interchanged. This will add a path difference, $L\frac{v^2}{c^2}$ in the opposite sense to that obtained before. Therefore one can observe a fringe-shift, $\Delta N = \frac{\text{Path differences}}{\text{Wavelength}} = \frac{2Lv^2}{\lambda c^2}$ where $c$ is the round-trip averaged speed of light. In the Michelson-Morley experiment in 1887, they used a light source $\lambda = 5000\text{Å}$, and path length of about 10 m. Therefore, their expected fringe-shift due to the velocity of the laboratory with respect to CMB ($\approx 300$ km/sec) should be, $\Delta N = 40$. However their measured fringe-shift using continuous rotation (from 0° to 90°) of the interferometer was zero [90].

During the half-century after 1881, there were about three dozen basic papers reporting the results of ether-drift experiments or of experiments closely related to Michelson's quest for evidence of relative motion of the earth through the ether [55]. Only D.C. Miller's experiments in the 1920s claimed to have found the long-sought absolute motion of the earth [56]. However, in 1955, the year that Albert Einstein died, Robert S. Shankland and his colleagues in Cleveland published an elaborate analysis of Miller's work, judging his anomalous, small but positive results to have been caused by inadequate temperature control [57]. A review of the classic Michelson-Morley experiments using RMS-test theory is presented in the Fig. 6.

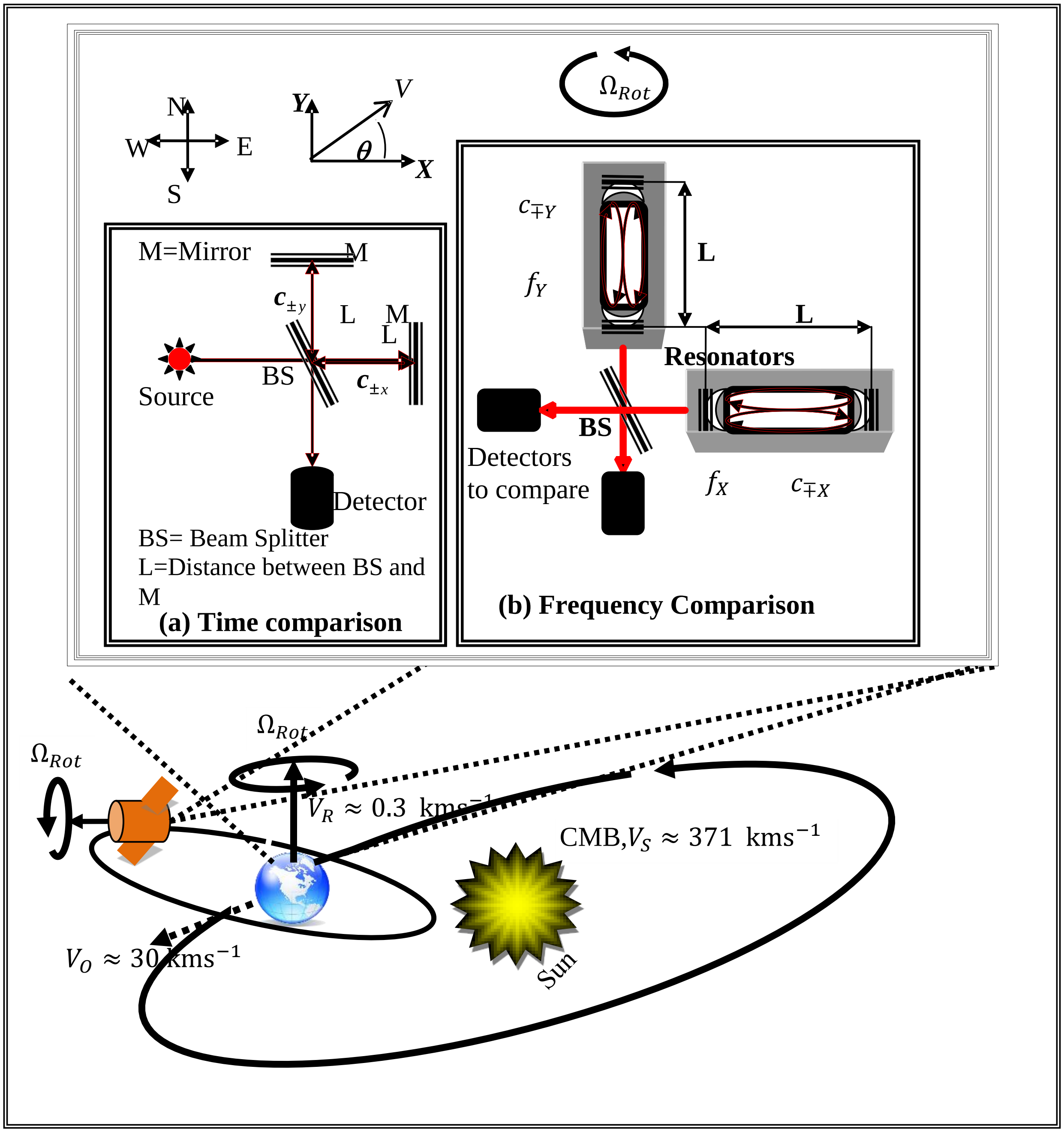

Figure 5. Schematic diagram of the Michelson-Morley (M-M) Experiments, (a) Classical (Interferometer), (b) Modern (Resonator), to test the isotropy of the speed of light since 1881. $\Omega_{Rot} = \frac{2\pi}{T}$, where $T$ =Earth's rotation rate (24 hrs) (or the turn table's rotation rate in lab) or the satellite's rotation rate (OPTIS: a satellite-based test [46]). $t_{\pm x}$ and $t_{\pm y}$ are times for round trip along X-axis for the light $c_{\pm x}$ and along Y-axis for the light $c_{\pm y}$ respectively. Also $f_X$ and $f_Y$ are frequencies for the resonators along X-axis and Y-axis respectively.

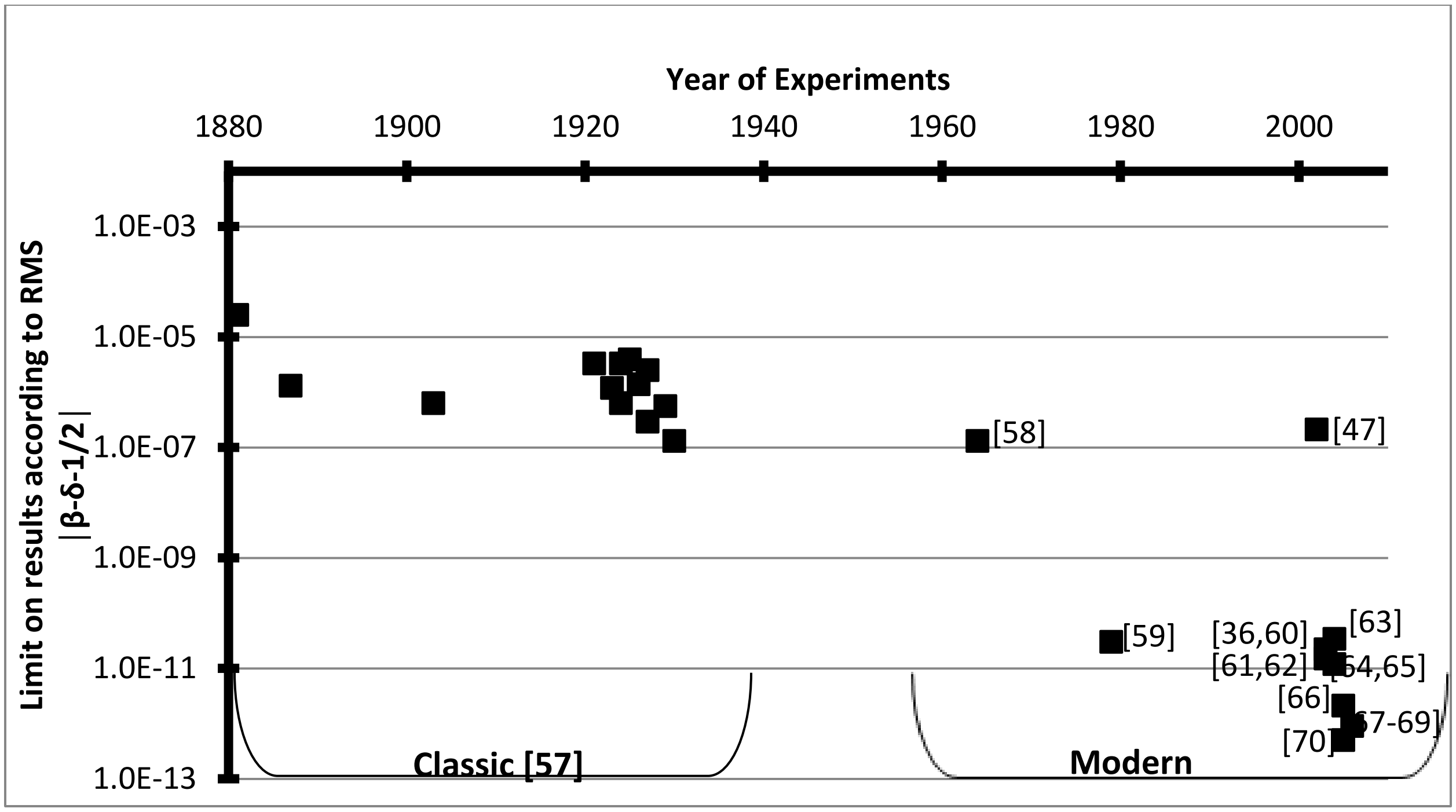

Figure 6. Selected experimental verifications (Classical to Modern) of the isotropy of the round-trip (two-way) velocity of light since 1881. The results are presented according to widely used Robertson-Mansouri-Sexl (RMS) test theory parameter $|\boldsymbol{\beta}-\boldsymbol{\delta}-\mathbf{1/2}|$. If Einstein's Special theory of relativity is valid then, $|\boldsymbol{\beta}-\boldsymbol{\delta}-\mathbf{1/2}|=\mathbf{0}$. Classical experiments are adopted from [57] and converted to RMS. Modern experiments are presented in Table-5.

### 3.1.2. Modern experiments :

A schematic of the modern Michelson-Morley experiment with different velocities of the laboratory is shown in Fig. 5(b). Jaseja *et al* in 1964 [58] performed the first modern type of Michelson-Morley experiment using resonators (cavities) as a sensitive test for an ether drift. They used the beams of two infrared lasers of slightly different frequency, combining by means of a beam splitter, and the resultant beat frequency was detected. This beat frequency is equal to the difference between the frequencies ($\Delta f = f_X - f_Y$ ) of the two laser beams. We know that any frequency of the laser can be written as $f = \frac{n}{2L}c$, where $n \in N$ and $c$ is the round-trip averaged velocity of the light inside the cavity. If both lasers are operated at about $3 \times 10^{14} Hz$, were rotated through 90°, the hypothetical variation of the speed of light should affect the frequencies of the lasers in the cavities and therefore a relative frequency change $\Delta f = 3$ MHz is expected at least from the orbital velocity of the earth, $30\ \text{kms}^{-1}$. But no change in the beat frequency was detected [58].

A review of the modern Michelson-Morley experiments using RMS-test theory is presented in the Fig. 6 and in the Table-5. The significant trials of the modern M-M experiment are summarized in the Table-5 where the first three columns recorded the observer, place, date and time of observations. The fourth column presents the rotation of the experiment- by the Earth or by a turn table. The fifth column gives the limit of the results in the RMS-framework. All these experiments presented in Fig. 6 are sensitive to measure the round-trip average speed of light and are comparable with ‘experiment (a)’ in Fig. 8.

## 3.2. The single-trip (one-way) experiments :

After the invention of masers, lasers and of the Mössabauer effect in around 1960, the one-way experiments became technically feasible. The Mössabauer effect: recoilless emission and absorption of gamma rays, has involvement with nuclear and electromagnetic interactions as well as the propagation of electromagnetic radiation, and is potentially a very powerful tool for one-way isotropy tests. The Mössabauer-rotor experiment is subject to relativistic time dilation where the dilatation factor can be deduced from the modified Doppler shift formula. One-way isotropy tests using Mössabauer-rotor experiments were performed by different observers in the 1960s [71 - 73]. A disk with a γ-ray emitter on the rim and an absorber at the centre where a detector was placed just behind the absorber was rotated. Observation of the directional dependence of the γ-rays transition through the absorber was monitored by the detector.

Recent reports by [11, 74 -76] present a series of measurements for the one-way isotropy of the speed of light tests performed at the GRAAL facility of the European Synchrotron Radiation Facility (ESRF) in Grenoble. These test the anisotropy of the speed of light by observing Compton scattering of laser photons on high-energy electrons. Zhou and Ma present a theoretical interpretation of the GRAAL one-way experiments in [11]. Also they present a brief review of some one-way and two-way experiments in their report. We make a comparison of reported limits of the one-way experiments in [11, 74 – 76] in Fig. 7.

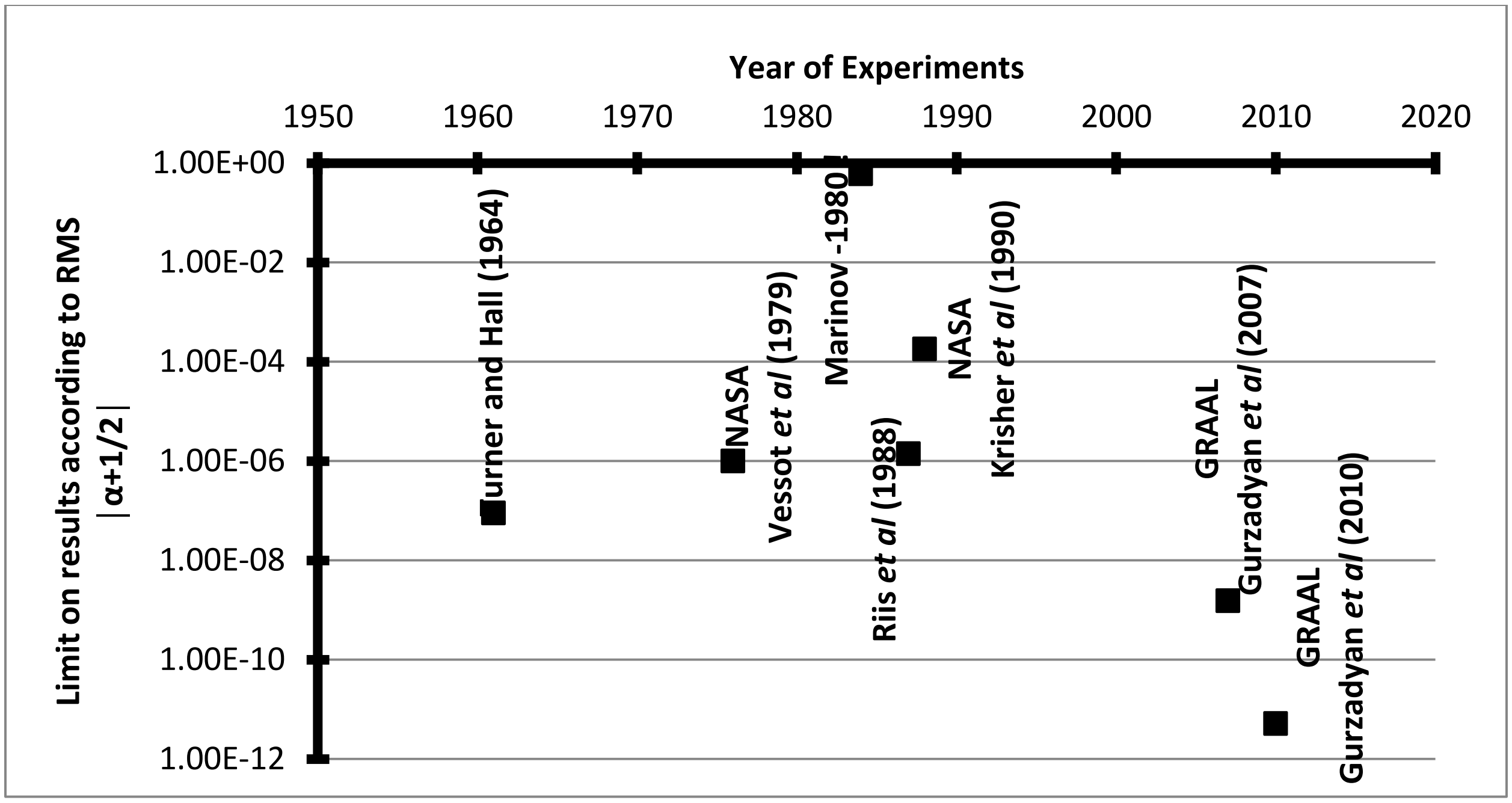

Figure 7. Selected experimental verifications (included the limits from [48] and [11] excluding Marinov-1880 [83] ) of the isotropy of the single-trip (one-way) velocity of light. The results are presented according to widely used Robertson-Mansouri-Sexl (RMS) test theory parameter $|\alpha+1/2|$. If Einstein's Special theory of relativity is valid then, $|\alpha+1/2|= 0$.

The Smithsonian Astrophysical Observatory – NASA Gravity Probe A (GP-A) Rocket Redshift experiment reported by Vessot *et al* [77, 78] which compared the rates of two hydrogen maser clocks one on the Scout rocket and other on the ground. The comparison as a function of the direction of the velocity of the rocket tests the isotropy of the one-way speed of light. The one-way experiment reported by Riis *et al* [49] and adopted by Will [48] compared the frequency of light emitted by atoms excited resonantly via two-photon absorption (TPA) in an atomic beam with the frequency of a stationary absorber as a function of the earth's rotation. This experiment was testing the isotropy of the first-order Doppler shift. Both experiments can be compared with experiment (b) in Fig. 8.

The JPL experiment was reported by Krisher *et al* [79] in 1990 and monitored the time-of-flight of light signals propagated in both directions along the fiber optic link between two hydrogen maser clocks. This experiment is compared in Fig. 8.

Marinov devoted himself to establishing the absoluteness of space-time by measuring the absolute velocity of the solar system by means of a coupled mirrors experiment [80], moving platform experiments [81] and coupled shutters experiment [82, 83]. These experiments were controversial as they disregarded STR. As shown by Duffy [84], the result of the coupled-mirrors experiment is a very different situation from his theory. Fizeau's gear-wheel method for measuring the speed of light [85] was adopted and improved by Marinov [82, 83] to measure the hypothetical variation of the one-way speed of light in what is called the coupled shutters experiment. This experiment was reported to detect significant light anisotropy but presented controversial results compared with other established results as shown in Fig.7 and in [86 - 88]. As we learn from previous experimental tests the pitfalls of temperature control are important concerns for the isotropy of light tests [55]. Observer's own body-heat or infrared radiation can produce an effect on the test results [89, 90]. Therefore, it is unclear how Marinov [82, 83] was controlling disturbances caused by local and temporal variations of temperature and other environmental disturbances such as pressure and humidity which could cause the effects of variation in the photo detectors' responses. Following [86] we would like to propose that it would be interesting to make an independent repetition of a gear-wheel type experiment performed by Marinov in any sophisticated laboratory. This experiment is under investigation in our Space Engineering laboratory at the Centre for Research in Earth & Space Science (CRESS), York University, Toronto, Canada. The independent improved design of this experiment, its mathematical interpretation and its result with extended periods of graphical representation will be publish in near future.

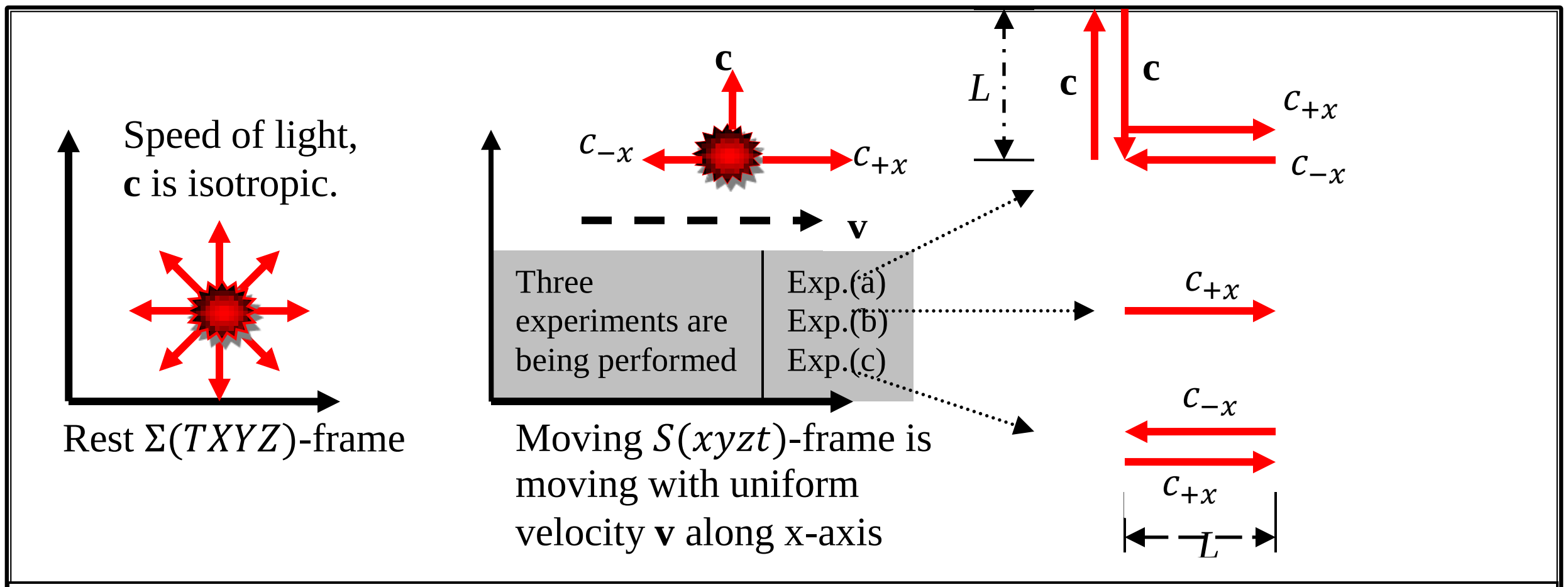

INTERPRETATION OF EXPERIMENTS (Using Galilean transformation as, $v \ll c$)

**Experiment (a)**

Following [91] time:

$$t_{+x} = \frac{L}{c_{+x}} = \frac{L}{c+v}$$

$$t_{-x} = \frac{L}{c_{-x}} = \frac{L}{c-v}$$

$$t_x = t_{+x} + t_{-x} = \frac{2Lc}{c^2 - v^2}$$

Similarly,

$$t_y = \frac{2L}{\sqrt{c^2 - v^2}}$$

$$\Delta t_a = t_x - t_y \cong \frac{Lv^2}{c^3}$$

**Experiment (b)**

Time,

$$t_{+x} = \frac{L}{c_{+x}} = \frac{L}{c+v}$$

Compared with a light in a rest frame.

Therefore,

$$t = \frac{L}{c}$$

$$\Delta t_b = t - t_{+x} = \frac{Lv}{c(c+v)}$$

**Experiment (c)**

Time,

$$t_{+x} = \frac{L}{c_{+x}} = \frac{L}{c+v}$$

$$t_{-x} = \frac{L}{c_{-x}} = \frac{L}{c-v}$$

Compared with opposite direction light signals in moving frame,

$$\Delta t_c = t_{-x} - t_{+x} = \frac{2Lv}{c^2 - v^2}$$

COMPARISON OF EXPERIMENTS [$v = 3 \times 10^5\ \mathrm{ms^{-1}}; c = 3 \times 10^8\ \mathrm{ms^{-1}}$]

$$\frac{\text{Experiment(a)}}{\text{Experiment(c)}} = \frac{\Delta t_a}{\Delta t_c} \cong \frac{v}{2c} = \frac{1}{2000}; \quad \frac{\text{Experiment(b)}}{\text{Experiment(c)}} = \frac{\Delta t_b}{\Delta t_c} \cong \frac{c-v}{2c} = \frac{1}{2}$$

**Experiment (a)**

(1) 2000 times less sensitive than Exp.(c).
(2) Question about the Lorentz contraction effect between two orthogonal arms.

**Experiment (b)**

(1) 2 times less sensitive than Exp.(c).
(2) No question about the Lorentz contraction effect.

**Experiment (c)**

(1) 2000 times more sensitive than Exp. (a) and 2 times more sensitive than Exp.(b).
(2) No question about the Lorentz contraction effect.

Figure 8. Comparison of experiments where $\Sigma(TXYZ)$ is a rest frame and $S(xyzt)$ is a moving frame with uniform velocity **v** along x-axis. If, Experiment (a) [comparable to the Michelson-Morley type experiments], Experiment (b) [comparable to the one-way speed of light for Doppler-shifted clock comparison] and Experiment (c) [comparable to two one-way speed of light comparison] are performed in the moving frame then the results can be compared as above.

Table 5. Modern Michelson-Morley tests of the isotropy of the velocity of light. All experiments are round-trip averaged speed of light measurements where frequencies are compared as shown in Fig. 5.

[$v = [300 \sim 377]$ kms$^{-1}$; $c = 3 \times 10^5$ kms$^{-1}$, RMS=Robertson-Mansouri-Sexl; SR= Special Relativity]

| **Observed by** | **Place** | **Date and Time of observation** | **Rotation by** | **Limits on Results According to RMS** $\left\|\beta - \delta - \frac{1}{2}\right\|$ **($= 0$ for SR)** |
|---|---|---|---|---|
| Jaseja *et al* 1964 [58] | New Bedford, USA | 01/20/63 (6 -12PM) | Turn-table | $\pm 1.3 \times 10^{-5}$ |
| Brillet & Hall 1979 [59] | Boulder, USA | 05/15 – 09/2, 1978 | Turn-table | $(3.0 \pm 4.9) \times 10^{-9}$ |
| Braxmaier *et al* 2002 [47] | Konstanz, Germany | 10/10/97( 190 days) | Earth | $\leq 2.1 \times 10^{-5}$ |
| Müller *et al* 2003 [36, 60] | Konstanz, Germany | 06/19/01-07/ 13/02 for 390 | Earth | $(2.2 \pm 1.5) \times 10^{-9}$ |
| Wolf *et al* 2002 -04 [61 - 65] | Paris France | 11/01-09/02 [60,61] | Earth | $(1.5 \pm 4.2) \times 10^{-9}$ [61,62] |
| | | 01/03-04/03 [62] | | $\leq 3.4 \times 10^{-9}$ [63] |
| | | 09/02-08/03 [63,64] | | $(1.2 \pm 2.2) \times 10^{-9}$ [64,65] |
| Herrmann *et al* 2005 [66] | Berlin, Germany | 12/04-04/05 | Turntable | $(2.1 \pm 1.9) \times 10^{-10}$ |
| Stanwix *et al* 2006 [67 - 69] | Crawley, Australia | 12/04-01/06 | Turntable | $(0.9 \pm 8.1) \times 10^{-10}$ |
| Antonini *et al* 2005 [70] | Düsseldorf | 02/04/05-02/08/05 | Turntable | $(0.5 \pm 3) \times 10^{-10}$ |

## 4. Discussion

From today's perspective the constancy of the speed of light influences a variety of areas from science-technology to philosophy [4 - 6]. Therefore to accept the idea of the constancy of the speed of light unambiguously, we need experiments sensitive enough to measure the hypothetical violation of the constancy of the speed of light. The Michelson-Morley experiment is beautiful in its simplicity, but tests only the constancy of the round-trip averaged speed of light. Based on the results of the classic or modern tests of Michelson-Morley experiment as shown in the in Fig. 6, we can only establish the special theory of relativity for the round-trip averaged speed of light. Also we note that Maxwell stated that no apparatus existed capable of measuring effects of the order $\left(\frac{v^2}{c^2}\right)$, the square of the ratio of the Earth's speed to that of the light [85].

In order to review isotropy tests of the single-trip (one-way) speed of light, we base our work on Table-1 of the article published by Will in [48] and Table 1 of the article published by Zhou and Ma in [11], and which are presented in Fig. 7 in this article. If we compare the one-way experiments of [48] in Fig. 7 with the two-way experiments in Fig. 6, the results are about 4 to 6 orders of magnitude smaller in the one-way experiments than those of two-way experiments. Also the most recent one-way experiment performed by Krisher *et al* [79] in 1988 in NASA- Jet Propulsion Laboratory Deep Space Network (DSN) presents 2 orders of magnitude smaller values than that of NASA's previous experiment by Vessot *et al* [77, 78] in 1976. This is contradictory to our expectation based on STR where we expect lower order of magnitude values with greater improvements. From 1976 to 1988, a twelve year period, science and technology improved and we expect more sensitive and accurate results. The results of the one-way experiments are increasing in magnitude with time, whereas, the two-way experiments are decreasing in magnitude with greater precision and improvements with time. However, the results from the limits of the one-way experiments of [11] at the GRAAL facility are consistent with STR. But the regularity in the variations of the reported results of the GRAAL measurements reported in [11] in different timeperiods remains unclear and needs further experimental investigations.

At extremely high energy levels the standard model of particle physics and Einstein's general theory of relativity theories coalesce into a single underlying unified theory where the prediction of the violation of the Lorentz invariance at a certain level demands more sensitive experimental tests [51]. We have presented a comparison of experiments in Fig. 8 that shows the one-way speed of light measurement is approximately 2000 times more sensitive than that of round-trip test. Will [48] showed that experiments which test the isotropy in one-way or two-way (round-trip) have observables that depend on test functions but not on the particular synchronization procedure. He noted that "the synchronization of clocks played no role in the interpretation of experiments provided that one is careful to express the results in terms of physically measurable quantities". Hence the synchronization is largely irrelevant and one-way speed of light is measurable. Therefore, we would like to propose that not only Michelson-Morley's two-way speed of light measurements be repeated but also other one-way speed of light measurements be performed with greater improvements. Results of the experimental tests spanning at least 24 hours periods in different seasons of the year should be recorded. Any hypothetical diurnal variations that might be observed should follow the figures presented in the section 2.2 in Fig. 3 and Fig. 4.

**Acknowledgements :** This work is supported in part by the National Science and Engineering Research Council, Thoth Technology Inc. and York University, Canada.